\documentclass[12pt]{article}

\usepackage{amssymb}
\usepackage{fullpage}
\usepackage{graphicx}
\usepackage{units}
\usepackage{fancybox}

\newtheorem{theorem}{Theorem}
\newtheorem{lemma}{Lemma}
\newtheorem{openproblem}{Open Problem}

\newcommand{\IE}{\mathbb{E}}
\newcommand{\CF}{\textsf{\sc CoinFlipper}}
\newcommand{\eps}{\varepsilon}

\newcommand{\YES}{\mathord{\it YES}}
\newcommand{\NO}{\mathord{\it NO}}
\newcommand{\TRUE}{\mathord{\it true}}

\newcommand{\qed}{\rule{0.5em}{1.5ex}}
\newcommand{\fqed}{{\hfill~\qed}}

\newcommand{\algAll}[2]{\vspace{0.5em}
  \begin{minipage}{.90\linewidth}%
    \shadowbox{%
      \begin{minipage}{\linewidth}%
        \textbf{#1}%
      \end{minipage}}
    \par
    {
      \fontencoding{OT1}\fontfamily{ppl}\small#2}
    \par\vspace{0.5em}
    \noindent\rule{\linewidth}{1mm} \linebreak
  \end{minipage}
  \vspace{0.5em}
}

\title{An Instance-Based Algorithm for Deciding the Bias of a Coin} 
\author{
Lu\'{i}s Fernando Schultz Xavier da 
Silveira\thanks{School of Electrical Engineering and Computer 
    Science, University of Ottawa, Canada. Supported by NSERC.} 
\and
Michiel Smid\thanks{School of Computer Science, 
                    Carleton University, Ottawa, Canada. 
                    Supported by NSERC.}} 
\date{\today}

\begin{document} 

\maketitle 

\begin{abstract} 
Let $q \in (0,1)$ and $\delta \in (0,1)$ be real numbers, and let $C$ be 
a coin that comes up heads with an unknown probability $p$, such 
that $p \neq q$. We present an algorithm that, on input $C$, $q$, and 
$\delta$, decides, with probability at least $1-\delta$, whether $p<q$ 
or $p>q$. The expected number of coin flips made by this algorithm is   
$O \left( \frac{\log\log(1/\eps) + \log(1/\delta)}{\eps^2} \right)$,
where $\eps = |p-q|$. 
\end{abstract} 

\section{Introduction} 
Let $q \in (0,1)$ and $\eps \in (0 , \min \{ q,1-q \} )$ be real 
numbers. Consider a coin that comes up heads with 
an unknown probability $p$ and, thus, comes up tails with probability 
$1-p$. Assume we know that $p \in \{ q+\eps , q-\eps \}$. 

Let $\delta \in (0,1)$ be a real number. The following algorithm 
decides, with probability at least $1 - \delta$, whether 
$p = q - \eps$ (this corresponds to the output $\YES$) or 
$p = q + \eps$ (this corresponds to the output $\NO$): 

\begin{itemize}
\item Flip the coin $k$ times, where 
\[ k = \left\lceil \frac{\ln (1/\delta)}{2 \eps^2} \right\rceil . 
\] 
\item Let $X$ be the number of heads in this sequence of $k$ coin flips. 
\begin{itemize} 
\item If $X \leq qk$, then return $\YES$. 
\item If $X > qk$, then return $\NO$. 
\end{itemize} 
\end{itemize} 

To prove correctness, assume first that $p = q - \eps$. By the 
Chernoff--Hoeffding bound (see Lemma~\ref{lemCH} below), we have 
\begin{eqnarray*} 
 \Pr(\, \mbox{the algorithm returns $\NO$}\, ) & = & 
            \Pr \left( X > qk \right) \\ 
  & \leq & \Pr \left( X \geq qk \right) \\ 
  & = & \Pr \left( X \geq pk + \eps k \right) \\
  & \leq & e^{-2 k \eps^2} \\
  & \leq & \delta 
\end{eqnarray*} 
and, therefore, with probability at least $1 - \delta$, the algorithm 
correctly returns $\YES$. By a symmetric argument, in case $p = q+\eps$, 
the algorithm correctly returns $\NO$ with probability at least 
$1 - \delta$. 

Observe that this algorithm must know the values of $q$, $\eps$, 
and $\delta$. The number of coin flips made by the algorithm is 
$O(\frac{\log(1/\delta)}{\eps^2})$, which is optimal for the case when 
$q=1/2$: Any algorithm that determines, with probability at least 
$1 - \delta$, whether $p = 1/2 - \eps$ or $p = 1/2 + \eps$, must flip 
the coin $\Omega(\frac{\log(1/\delta)}{\eps^2})$ times in the worst 
case. For a proof of this claim, see Lemma~5.1 in Anthony and 
Bartlett~\cite{ab-nnl-99}. The results by Mannor and 
Tsitsiklis~\cite{st-04} imply the same lower bound for the expected 
number of coin flips made by any algorithm that uses, besides flipping 
the coin, randomization to decide when to terminate. 

In this paper, we consider a more general version of this problem. 
Besides the coin having an unknown probability $p \in (0,1)$ of coming 
up heads, we are given a real number $q \in (0,1)$ such that $p \neq q$
and a real number $\delta \in (0,1)$. How can we decide whether $p$ is 
smaller than or larger than $q$? 

More formally, we consider the problem of designing an algorithm that 
takes as input the above coin and the real numbers $q \in (0,1)$ and 
$\delta \in (0,1)$, and outputs $\YES$ or $\NO$, such that 
\begin{enumerate} 
\item if $p<q$, then the output is $\YES$ with probability at 
least $1 - \delta$, 
\item if $p>q$, then the output is $\NO$ with probability at 
least $1 - \delta$. 
\end{enumerate} 
Any such algorithm will repeatedly flip the coin and determine its output 
based on the resulting sequence of heads and tails. The goal is to 
minimize the number of coin flips made by the algorithm. Intuitively, 
this number should depend on the absolute value $|p-q|$ of the difference 
between $p$ and $q$: The smaller this value is, the more coin flips are 
needed to decide which of $p$ and $q$ is larger.  

An obvious approach is as follows. For a given value $\eps \in (0,1)$,
the algorithm flips the coin $k$ times, where 
\begin{equation}  \label{eqK}  
   k = \left\lceil \frac{\ln(1/\delta)}{2 \eps^2} \right\rceil . 
\end{equation} 
Let $X$ be the number of heads in this sequence of coin flips. The 
Chernoff--Hoeffding bound implies the following two claims: If $p<q$, 
then 
\[ \Pr \left( X \geq qk + \eps k \right) \leq 
   \Pr \left( X \geq pk + \eps k \right) \leq 
   e^{-2k \eps^2} \leq \delta . 
\]
If $p>q$, then 
\[ \Pr \left( X \leq qk - \eps k \right) \leq 
   \Pr \left( X \leq pk - \eps k \right)  \leq 
   e^{-2k \eps^2} \leq \delta . 
\]
Based on the value of $X$, the algorithm does the following:   
\begin{itemize} 
\item If $X \leq qk - \eps k$, it returns $\YES$. 
\item if $X \geq qk + \eps k$, it returns $\NO$. 
\item Otherwise, the algorithm does not have enough information to 
      decide which of $p$ and $q$ is larger. In this case, the algorithm
      chooses a smaller value of $\eps$, recomputes the value of $k$ 
      according to (\ref{eqK}), and repeats. 
\end{itemize} 

A natural choice for the values of $\eps$ is the sequence $1/2^i$ for 
$i=1,2,3,\ldots$. Since each iteration of this algorithm depends on the 
outcomes of all previous iterations, it is not clear that this algorithm 
is correct with probability at least $1 - \delta$. 

In this paper, we show that we do obtain a correct algorithm, if we take 
a slightly larger value for $k$: In (\ref{eqK}), we replace 
$\ln (1/\delta)$ by $\ln(\pi^2 i^2/(6\delta))$.  

Let 
\[ d = \left\lceil \log \left( \frac{1}{|p-q|} \right) \right\rceil , 
\]
where $\log$ is the logarithm to the base $2$. If, for example, $q=1/2$ 
and $p>q$, then $d$ is the position of the leftmost bit in which the 
(infinite) binary representations of $p$ and $q$ differ. 
We can think of $d$ as being the degree of ``difficulty'': The larger  
$d$ is, the closer $p$ and $q$ are to each other and, thus, the more 
``difficult'' it is to decide whether $p<q$ or $p>q$. 

Using the new value for $k$, we prove the following: 
\begin{enumerate} 
\item The output of the algorithm is correct with probability at 
      least $1 - \delta$. 
\item The expected number of iterations made by the algorithm is at 
      most 
      \[ d + 1.2 = \log \left( \frac{1}{|p-q|} \right) + O(1) . 
      \] 
\item The expected total number of coin flips made by the algorithm 
      and its expected running time are 
      \[ O \left( 4^d \cdot \log (d/\delta) \right) = 
         O \left( \left( \log\log \left( \frac{1}{|p-q|} \right) 
                        + \log (1/\delta) 
                  \right) \cdot \frac{1}{(p-q)^2} 
           \right) . 
      \] 
\end{enumerate} 

\section{The Algorithm} 
Below, we give a formal description of the algorithm. In 
Section~\ref{secAN}, we will analyze the success probability, the 
expected number of iterations, and the expected total number of coin 
flips. 

\algAll{Algorithm $\CF(C,q,\delta)$}{
{\bf Comment:} $C$ is a coin with an unknown probability of coming 
up heads, and $q \in (0,1)$ and $\delta \in (0,1)$ are real numbers.
All coin flips are mutually independent. 

\begin{quote}
\begin{tabbing}
$i=1$; \\
{\bf while} $\TRUE$ \\ 
{\bf do} \= $\eps = 1/2^i$; \\ 
         \> $k = \lceil \ln \left( \frac{\pi^2 i^2}{6 \delta} \right) 
                         / \left(2 \eps^2 \right) 
                 \rceil$; \\ 
         \> flip the coin $k$ times; \\
         \> $X=$ number of heads; \\ 
         \> {\bf if} $X \leq qk - \eps k$ \\ 
         \> {\bf then} return $\YES$ and terminate \\ 
         \> {\bf else} \= {\bf if} $X \geq qk + \eps k$ \\ 
         \>            \> {\bf then} return $\NO$ and terminate \\
         \>            \> {\bf else} $i=i+1$ \\ 
         \>            \> {\bf endif} \\
         \> {\bf endif} \\ 
{\bf endwhile}  
\end{tabbing}
\end{quote}
}

\section{The Analysis of Algorithm $\CF$}
\label{secAN}  
 
Our analysis will use the additive version of the well known 
Chernoff--Hoeffding bound (see, e.g., Theorem~1.1 
in Dubhashi and Panconesi~\cite{dp-com-09}):  

\begin{lemma}[Chernoff--Hoeffding] \label{lemCH} 
Let $k \geq 1$ be an integer and let $p \in (0,1)$ be a real number. 
Consider a coin that comes up heads with probability $p$. Let $X$ be the 
random variable that counts the number of heads in a sequence of $k$ 
mutually independent coin flips. Then, for any real number 
$\eps \in (0,1)$, 
\[ \Pr \left( X \geq pk + \eps k \right) \leq 
      e^{-2k \eps^2} 
\]
and  
\[ \Pr \left( X \leq pk - \eps k \right) \leq 
      e^{-2k \eps^2} .
\]
\end{lemma} 

Throughout the rest of this section, $p$ denotes the (unknown) probability 
that the coin $C$ comes up heads, and $q$ and $\delta$ are the real 
numbers that are the input to algorithm $\CF(C,q,\delta)$. 
We will assume throughout that $p>q$. The analysis for the case when 
$p<q$ is symmetric. 

\subsection{The Success Probability} 
We have to prove that, with probability at least $1 - \delta$, 
algorithm $\CF(C,q,\delta)$ returns $\NO$. Thus, if we let $A$ be the 
event 
\[ A = \mbox{``algorithm $\CF(C,q,\delta)$ returns $\YES$'',} 
\] 
then we have to prove that $\Pr(A) \leq \delta$. 

For each integer $i \geq 1$, define the events 
\begin{eqnarray*} 
  A_i & = & \mbox{ ``algorithm $\CF(C,q,\delta)$ returns $\YES$ in 
                     iteration $i$'',} \\
  B_i & = & \mbox{ ``iteration $i$ of algorithm $\CF(C,q,\delta)$ 
                     takes place''.} 
\end{eqnarray*}
Observe that the events $A_i$ and $A_i \cap B_i$ are the same. Using 
this, together with the fact that the event $A$ is the pairwise disjoint 
union of the $A_i$'s, we have 
\begin{eqnarray*} 
 \Pr(A) & = & \Pr \left( \bigcup_{i=1}^{\infty} A_i \right) \\  
  & = & \sum_{i=1}^{\infty} \Pr \left( A_i \right) \\ 
  & = & \sum_{i=1}^{\infty} \Pr \left( A_i \cap B_i \right) \\ 
  & = & \sum_{i=1}^{\infty} 
           \Pr \left( A_i \mid B_i \right) \cdot 
           \Pr \left( B_i \right) \\ 
  & \leq & \sum_{i=1}^{\infty} \Pr \left( A_i \mid B_i \right) . 
\end{eqnarray*} 

Let $i \geq 1$ be an integer. We will derive an upper bound on 
$\Pr \left( A_i \mid B_i \right)$. Consider iteration~$i$ of 
algorithm $\CF(C,q,\delta)$, and the values of $\eps$, $k$, and 
$X$ during this iteration. It follows from the algorithm that 
\begin{eqnarray*} 
  \Pr \left( A_i \mid B_i \right) & = & 
  \Pr \left( X \leq q k - \eps k \right) \\ 
   & \leq & \Pr \left( X \leq p k - \eps k \right) ,  
\end{eqnarray*} 
where the inequality follows from the assumption that $p>q$. 
Lemma~\ref{lemCH} implies that 
\[ \Pr \left( A_i \mid B_i \right) \leq e^{-2k \eps^2} .
\] 
Since 
\[ 2 k \eps^2 \geq  
      \ln \left( \frac{\pi^2 i^2}{6 \delta} \right) ,
\] 
it follows that 
\[ \Pr \left( A_i \mid B_i \right) \leq 
     \frac{6 \delta}{\pi^2} \cdot \frac{1}{i^2} . 
\]

Using the well known identity $\sum_{i=1}^{\infty} 1/i^2 = \pi^2 / 6$, 
we conclude that 
\begin{eqnarray*} 
 \Pr(A) & \leq & 
   \frac{6 \delta}{\pi^2} \sum_{i=1}^{\infty} \frac{1}{i^2} \\ 
 & = & \frac{6 \delta}{\pi^2} \cdot \frac{\pi^2}{6} \\ 
 & = & \delta . 
\end{eqnarray*} 

\subsection{The Expected Number of Iterations} 
\label{subsecIT}
Let $Y$ be the random variable that counts the number of iterations made 
when running algorithm $\CF(C,q,\delta)$. For each integer $i \geq 1$, 
define the indicator random variable 
\[ Y_i = 
   \left\{ \begin{array}{ll} 
            1 & \mbox{if iteration $i$ takes place,} \\ 
            0 & \mbox{otherwise.} 
           \end{array} 
   \right. 
\] 
Then 
\[ Y = \sum_{i=1}^{\infty} Y_i . 
\] 
Let 
\[ d = \left\lceil \log \left( \frac{1}{|p-q|} \right) \right\rceil . 
\]
Observe that $d \geq 1$ and, because of our assumption that $p>q$,  
\[ q + 1/2^d \leq p < q + 1/2^{d-1} .
\] 
Using the Linearity of Expectation, we have 
\begin{eqnarray*} 
 \IE(Y) & = & \IE \left( \sum_{i=1}^{\infty} Y_i \right) \\ 
  & = & \sum_{i=1}^{d+1} \IE \left( Y_i \right) + 
        \sum_{j=1}^{\infty} \IE \left( Y_{d+j+1} \right) \\ 
  & \leq & d+1 + \sum_{j=1}^{\infty} \IE \left( Y_{d+j+1} \right) .  
\end{eqnarray*} 

Let $j \geq 1$ be an integer. Consider iteration $d+j$ of algorithm 
$\CF(C,q,\delta)$, and the values of $\eps$, $k$, and $X$ during this 
iteration. We have 
\begin{eqnarray*} 
  \IE \left( Y_{d+j+1} \right) & = & \Pr \left( B_{d+j+1} \right) \\ 
  & = & \Pr \left( B_{d+j+1} \cap B_{d+j} \right) \\ 
  & = & \Pr \left( B_{d+j+1} \mid B_{d+j} \right) \cdot 
          \Pr \left( B_{d+j} \right) \\ 
  & \leq & \Pr \left( B_{d+j+1} \mid B_{d+j} \right) \\ 
  & = & \Pr \left( qk - \eps k < X < qk + \eps k \right) \\ 
  & \leq & \Pr \left( X < qk + \eps k \right) .  
\end{eqnarray*} 
Since $q \leq p - 1/2^d$, we have 
\begin{eqnarray*} 
  \IE \left( Y_{d+j+1} \right) & \leq & 
  \Pr \left( X \leq \left( p -1/2^d \right) k + \eps k \right) \\ 
  & = & \Pr \left( X \leq pk - \left( 1/2^d - \eps \right) k \right) . 
\end{eqnarray*} 
Observe that 
\[ 1/2^d - \eps = 1/2^d - 1/2^{d+j} \geq 1/2^d - 1/2^{d+1} = 1/2^{d+1} , 
\] 
implying that 
\[ \IE \left( Y_{d+j+1} \right) \leq 
  \Pr \left( X \leq pk - k/2^{d+1} \right) . 
\] 
Using Lemma~\ref{lemCH}, we obtain 
\[ \IE \left( Y_{d+j+1} \right) \leq e^{-k/2^{2d+1}} . 
\]
It follows from the algorithm that 
\begin{eqnarray*} 
 \frac{k}{2^{2d+1}} & \geq &  
 \frac{\ln \left( \frac{\pi^2 (d+j)^2}{6 \delta} \right)}{2 \eps^2} 
   \cdot \frac{1}{2^{2d+1}} \\ 
 & \geq & 
 \frac{\ln \left( \frac{4 \pi^2}{6 \delta} \right)}{2 \eps^2} 
   \cdot \frac{1}{2^{2d+1}} \\ 
 & \geq & 
 \frac{\ln \left( 6 / \delta \right)}{2 \eps^2} 
   \cdot \frac{1}{2^{2d+1}} \\ 
 & = & 4^{j-1} \cdot \ln \left( 6/\delta \right) \\
 & \geq & 4^{j-1} \cdot \ln 6 . 
\end{eqnarray*} 
Therefore, 
\begin{equation}   \label{eqIE} 
  \IE \left( Y_{d+j+1} \right) \leq (1/6)^{4^{j-1}} . 
\end{equation} 
Thus, 
\[ \IE(Y) \leq d+1 + \sum_{j=1}^{\infty} (1/6)^{4^{j-1}} .
\]
The infinite series converges and its value is approximately  
$0.167438$, which is less than $0.2$. We conclude that 
\begin{eqnarray*} 
 \IE(Y) & \leq & d + 1.2 \\ 
  & = & \left\lceil \log \left( \frac{1}{|p-q|} \right) 
        \right\rceil + 1.2 . 
\end{eqnarray*}

\subsection{The Expected Total Number of Coin Flips} 
Let $Z$ be the random variable that counts the total number of coin 
flips made when running algorithm $\CF(C,q,\delta)$.  
Using the indicator random variables $Y_i$ of Section~\ref{subsecIT},
and denoting the value of $k$ in iteration $i$ by $k_i$, we have 
\begin{eqnarray} 
 \IE(Z) & = & \IE \left( \sum_{i=1}^{\infty} Y_i \cdot k_i \right) 
            \nonumber \\ 
 & = & \sum_{i=1}^{d+1} \IE \left( Y_i \right) \cdot k_i + 
       \sum_{j=1}^{\infty} \IE \left( Y_{d+j+1} \right) \cdot k_{d+j+1} 
            \nonumber \\  
 & \leq & \sum_{i=1}^{d+1} k_i + 
       \sum_{j=1}^{\infty} \IE \left( Y_{d+j+1} \right) \cdot k_{d+j+1} . 
          \label{eqhello}  
\end{eqnarray} 

Since, for $1 \leq i \leq d+1$,  
\begin{eqnarray*} 
   k_i & \leq & 1 + \frac{1}{2} \cdot 4^i \cdot 
               \ln \left( \frac{\pi^2 i^2}{6 \delta} \right) \\ 
    & \leq & 1 + \frac{1}{2} \cdot 4^i \cdot 
               \ln \left( \frac{\pi^2 (d+1)^2}{6 \delta} \right) , 
\end{eqnarray*} 
we obtain the following upper bound on the first summation in 
(\ref{eqhello}): 
\begin{eqnarray} 
 \sum_{i=1}^{d+1} k_i & \leq & d+1 + 
 \frac{1}{2} \cdot \ln \left( \frac{\pi^2 (d+1)^2}{6 \delta} \right) 
  \sum_{i=1}^{d+1} 4^i \nonumber \\ 
 & = & O \left( 4^d \cdot \log (d/\delta) \right) \label{eq7} .  
\end{eqnarray} 

To bound the second summation in (\ref{eqhello}), let $j \geq 1$. 
Using (\ref{eqIE}), we have  
\[ \IE \left( Y_{d+j+1} \right) \cdot k_{d+j+1} \leq 
    \left( \frac{1}{6} \right)^{4^{j-1}} 
     \left( 1 + \frac{1}{2} \cdot 4^{d+j+1} \cdot 
               \ln \left( \frac{\pi^2 (d+j+1)^2}{6 \delta} \right) 
     \right) . 
\] 
Since $d+j+1 \leq 3dj$, it follows that 
\begin{eqnarray*} 
  \ln \left( \frac{\pi^2 (d+j+1)^2}{6 \delta} \right) & \leq & 
  \ln \left( \frac{9 \pi^2 (dj)^2}{6 \delta} \right) \\ 
 & \leq & \ln \left( \frac{3 \pi^2}{2} \right) + 
          \ln \left( \frac{(dj)^2}{\delta^2} \right) \\ 
 & \leq & 3 + 2 \cdot \ln \left( \frac{dj}{\delta} \right) . 
\end{eqnarray*} 
Thus, we obtain the following upper bound on the second summation in 
(\ref{eqhello}): 
\begin{eqnarray} 
  \sum_{j=1}^{\infty} \IE \left( Y_{d+j+1} \right) \cdot k_{d+j+1} 
   & \leq & 
  \sum_{j=1}^{\infty} 
    \left( \frac{1}{6} \right)^{4^{j-1}} 
     \left( 1 + \frac{1}{2} \cdot 4^{d+j+1} \cdot 
               \left( 3 + 2 \cdot \ln \left( \frac{dj}{\delta} \right) 
               \right) 
     \right) \nonumber \\  
 & = & \sum_{j=1}^{\infty} \left( \frac{1}{6} \right)^{4^{j-1}} + 
               \label{sum1} \\ 
 &   & 4^{d+1} \left( \frac{3}{2} + \ln \left( \frac{d}{\delta} \right) 
               \right) 
       \sum_{j=1}^{\infty} 
          4^j \cdot \left( \frac{1}{6} \right)^{4^{j-1}} + 
                \label{sum2} \\ 
 &   & 4^{d+1} 
       \sum_{j=2}^{\infty} 
          4^j \cdot \left( \frac{1}{6} \right)^{4^{j-1}} \ln j . 
              \label{sum3}  
\end{eqnarray} 
The infinite series in (\ref{sum1}), (\ref{sum2}), and (\ref{sum3}) 
converge, and their values are approximately $0.167438$, $0.679012$, 
and $0.00855737$, respectively.   
Thus, 
\begin{equation}   \label{eq6}
   \sum_{j=1}^{\infty} \IE \left( Y_{d+j+1} \right) \cdot k_{d+j+1} 
   = O \left( 4^d \cdot \log (d/\delta) \right) .  
\end{equation}  

By combining (\ref{eqhello}), (\ref{eq7}) and (\ref{eq6}), we obtain our 
upper bound on the expected total number of coin flips made by 
algorithm $\CF(C,q,\delta)$:  
\begin{eqnarray*} 
  \IE(Z) & = & O \left( 4^d \cdot \log (d/\delta) \right) \\ 
  & = & O \left( \left( \log\log \left( \frac{1}{|p-q|} \right) 
                        + \log (1/\delta) 
                 \right) \cdot \frac{1}{(p-q)^2} 
          \right) . 
\end{eqnarray*} 
Since the running time of algorithm $\CF(C,q,\delta)$ is proportional 
to $Z$, we obtain the same upper bound on its expected running time. 

The following theorem summarizes our result. 

\begin{theorem}  \label{thm1} 
Let $q \in (0,1)$ and $\delta \in (0,1)$ be real numbers, and let $C$ be 
a coin that comes up heads with an unknown probability $p$, such 
that $p \neq q$. Algorithm $\CF(C,q,\delta)$ has the following 
properties: 
\begin{enumerate} 
\item If $p<q$, then the output is $\YES$ with probability at least 
      $1 - \delta$. 
\item If $p>q$, then the output is $\NO$ with probability at least 
      $1 - \delta$. 
\item Let $\eps = |p-q|$. The expected total number of coin flips is 
      \[ O \left( \frac{\log\log(1/\eps) + \log(1/\delta)}{\eps^2} 
           \right) .
      \] 
\end{enumerate} 
\end{theorem} 

\begin{openproblem} 
The expected number of coin flips in Theorem~\ref{thm1} contains the 
term $\log\log(1/\eps)$, where the value of $\eps$ is not known when 
the algorithm starts. Does there exist an algorithm that solves the 
problem by making, in expectation, 
$O \left( \frac{\log(1/\delta)}{\eps^2} \right)$ coin flips? 
\end{openproblem}

\bibliographystyle{plain}
\bibliography{flipcoin}

\begin{thebibliography}{1}

\bibitem{ab-nnl-99}
M.~Anthony and P.~L. Bartlett.
\newblock {\em Neural Network Learning: Theoretical Foundations}.
\newblock Cambridge University Press, Cambridge, UK, 1999.

\bibitem{dp-com-09}
D.~P. Dubhashi and A.~Panconesi.
\newblock {\em Concentration of Measure for the Analysis of Randomized
  Algorithms}.
\newblock Cambridge University Press, Cambridge, UK, 2009.

\bibitem{st-04}
S.~Mannor and J.~N. Tsitsiklis.
\newblock The sample complexity of exploration in the multi-armed bandit
  problem.
\newblock {\em Journal of Machine Learning Research}, 5:623--648, 2004.

\end{thebibliography}

\end{document}